\documentclass[a4paper,11pt]{article}

\usepackage{blkarray}
\usepackage[hyphens]{url}
\usepackage{mathtools,amssymb}
\usepackage[usenames,dvipsnames,table]{xcolor}
\usepackage[titletoc]{appendix}
\usepackage{mathtools,amssymb}
\usepackage[makeroom]{cancel}
\usepackage{graphicx}
\usepackage{epstopdf}
\usepackage{mathrsfs}
\usepackage{enumitem}
\usepackage{stackrel}
\usepackage{float}
\usepackage{booktabs}
\usepackage[normalem]{ulem}  
\usepackage{cancel} 
\usepackage{hyperref}
\usepackage[final]{showkeys} 
\usepackage{longtable}
\usepackage{physics}
\usepackage{empheq}
\usepackage{cite} 
\usepackage{comment}
\usepackage{tikz}
\usetikzlibrary{decorations.pathmorphing}
\usetikzlibrary{decorations.pathreplacing}
\usetikzlibrary{decorations.text}
\usepackage{braket}
\usepackage{fullpage}
\usepackage{subcaption}
\usepackage{pgfplots}
\usepackage{amsmath}
\usepackage{ytableau}

\usepackage{listings}
\usepackage{mathtools,amssymb}
\usepackage[usenames,dvipsnames,table]{xcolor}
\usepackage[titletoc]{appendix}
\usepackage{graphicx}
\usepackage{mathrsfs}
\usepackage{enumitem} 
\usepackage{float} 
\usepackage{booktabs} 
\usepackage[normalem]{ulem}  
\usepackage{cancel} 
\usepackage{longtable} 
\usepackage{physics} 
\usepackage{empheq} 
\usepackage{cite} 
\usepackage{comment}
\usepackage{braket}
\usepackage{fullpage}
\usepackage{tikz}
\usetikzlibrary{decorations.pathreplacing}
\usepackage[compat=1.1.0]{tikz-feynman}

\DeclareMathAlphabet{\mathpzc}{OT1}{pzc}{m}{it}
\DeclareMathSymbol{\mlq}{\mathord}{operators}{``}
\DeclareMathSymbol{\mrq}{\mathord}{operators}{`'}

\def\nn{\nonumber\\}
\def\psid{\psi^\dagger}

\def\ad{a^\dagger}

\allowdisplaybreaks
\allowbreak

\let\equation\gather
\let\endequation\endgather
\definecolor{refkey}{gray}{0.75}
\definecolor{labelkey}{RGB}{155,48,48}
\renewcommand*\showkeyslabelformat[1]{%
	\fbox{\parbox[t]{0.8\marginparwidth}{\raggedright\normalfont\scriptsize\url{#1}}}}

\usepackage{etoolbox}
\makeatletter
\patchcmd{\hyper@makecurrent}{%
	\ifx\Hy@param\Hy@chapterstring
	\let\Hy@param\Hy@chapapp
	\fi
}{%
	\iftoggle{inappendix}{
		\@checkappendixparam{chapter}%
		\@checkappendixparam{section}%
		\@checkappendixparam{subsection}%
		\@checkappendixparam{subsubsection}%
		\@checkappendixparam{paragraph}%
		\@checkappendixparam{subparagraph}%
	}{}%
}{}{ \errmessage{failed to patch}}

\newcommand*{\@checkappendixparam}[1]{%
	\def\@checkappendixparamtmp{#1}%
	\ifx\Hy@param\@checkappendixparamtmp
	\let\Hy@param\Hy@appendixstring
	\fi
}
\makeatletter

\newtoggle{inappendix}
\togglefalse{inappendix}

\apptocmd{\appendix}{\toggletrue{inappendix}}{}{\errmessage{failed to patch}}
\apptocmd{\subappendices}{\toggletrue{inappendix}}{}{\errmessage{failed to patch}}

\newcommand{\lsim}{\mathrel{\hbox{\rlap{\lower .55ex
				\hbox{$\sim$}} \kern-.3em \raise.4ex \hbox{$<$}}}}
\newcommand{\gsim}{\mathrel{\hbox{\rlap{\lower.55ex
				\hbox{$\sim$}} \kern-.3em \raise.4ex \hbox{$<$}}}}

\begin{document}
	
	
\newcommand{\partiald}[2]{\dfrac{\partial #1}{\partial #2}}
\newcommand{\be}{\begin{equation}}
\newcommand{\ee}{\end{equation}}
\newcommand{\f}{\frac}
\newcommand{\s}{\sqrt}
\newcommand{\lm}{\mathcal{L}}
\newcommand{\wm}{\mathcal{W}}
\newcommand{\om}{\omega}
\newcommand{\bea}{\begin{eqnarray}}
\newcommand{\eea}{\end{eqnarray}}
\newcommand{\ba}{\begin{align}}
\newcommand{\ea}{\end{align}}
\newcommand{\ep}{\epsilon}
\newcommand{\h}{\hat}
\def\ad{{a^\dagger}}
\def\psid{{\psi^\dagger}}
\def\ads{AdS$_{\text{2}}$~}
\def\gap#1{\vspace{#1 ex}}
\def\be{\begin{equation}}
\def\ee{\end{equation}}
\def\bal{\begin{array}{l}}
\def\ba#1{\begin{array}{#1}}  
\def\ea{\end{array}}
\def\bea{\begin{eqnarray}}
\def\eea{\end{eqnarray}}
\def\beas{\begin{eqnarray*}}
\def\eeas{\end{eqnarray*}}
\def\del{\partial}
\def\eq#1{(\ref{#1})}
\def\fig#1{Fig \ref{#1}} 
\def\re#1{{\bf #1}}
\def\bull{$\bullet$}
\def\nn{\nonumber}
\def\ub{\underbar}
\def\nl{\hfill\break}
\def\ni{\noindent}
\def\bibi{\bibitem}
\def\vev#1{\langle #1 \rangle} 
\def\mattwo#1#2#3#4{\left(\begin{array}{cc}#1&#2\\#3&#4\end{array}\right)} 
\def\tgen#1{T^{#1}}
\def\half{\frac12}
\def\floor#1{{\lfloor #1 \rfloor}}
\def\ceil#1{{\lceil #1 \rceil}}
	
\def\Tr{{\rm Tr}}
		
\def\mysec#1{\gap1\ni{\bf #1}\gap1}
\def\mycap#1{\begin{quote}{\footnotesize #1}\end{quote}}
		
\def\Red#1{{\color{red}#1}}
\def\blue#1{{\color{blue}#1}}
\def\Om{\Omega}
\def\a{\alpha}
\def\b{\beta}
\def\l{\lambda}
\def\g{\gamma}
\def\ep{\epsilon}
\def\Si{\Sigma}
\def\p{\phi}
\def\z{\zeta}

\def\lan{\langle}
\def\ran{\rangle}

\def\bit{\begin{item}}
\def\eit{\end{item}}
\def\benu{\begin{enumerate}}
\def\eenu{\end{enumerate}}
\def\fr#1#2{{\frac{#1}{#2}}}
\def\gsq{{{\tilde g}^2}}
	
\def\tr{{\rm tr}}
\def\intk#1{{\int\kern-#1pt}}


\newcommand{\com}{\textcolor{red}}
\newcommand{\new}[1]{{\color[rgb]{1.0,0.,0}#1}}
\newcommand{\old}[1]{{\color[rgb]{0.7,0,0.7}\sout{#1}}}
		
\renewcommand{\real}{\ensuremath{\mathbb{R}}}
		
\newcommand*{\Cdot}[1][1.25]{%
	\mathpalette{\CdotAux{#1}}\cdot%
		}
		\newdimen\CdotAxis
		\newcommand*{\CdotAux}[3]{%
			{%
				\settoheight\CdotAxis{$#2\vcenter{}$}%
				\sbox0{%
					\raisebox\CdotAxis{%
						\scalebox{#1}{%
							\raisebox{-\CdotAxis}{%
								$\mathsurround=0pt #2#3$%
							}%
						}%
					}%
				}%
				\dp0=0pt %
				\sbox2{$#2\bullet$}%
				\ifdim\ht2<\ht0 %
				\ht0=\ht2 %
				\fi
				\sbox2{$\mathsurround=0pt #2#3$}%
				\hbox to \wd2{\hss\usebox{0}\hss}%
			}%
		}
		
\newcommand\hcancel[2][black]{\setbox0=\hbox{$#2$}%
\rlap{\raisebox{.45\ht0}{\textcolor{#1}{\rule{\wd0}{1pt}}}}#2} 
		
\renewcommand{\arraystretch}{2.5}%
\renewcommand{\floatpagefraction}{.8}%
	
\def\newthing{\marginpar{{\color{red}****}}}
\def\doubt{{\color{red}{***}}}
\reversemarginpar
		
		
\def\tu{\tau}
\def\ze{z}
\def\d{\partial}
\def\L{\varphi}  
		
\DeclareRobustCommand{\rchi}{{\mathpalette\irchi\relax}}
\newcommand{\irchi}[2]{\raisebox{\depth}{$#1\chi$}}

\tikzfeynmanset{compat=1.0.0}
\allowdisplaybreaks
\allowbreak
\let\equation\gather
\let\endequation\endgather
\definecolor{refkey}{rgb}{0.9451,0.2706,0.4941}
\definecolor{labelkey}{rgb}{0.9451,0.2706,0.4941}
\renewcommand*\showkeyslabelformat[1]{\fbox{\parbox[t]{0.8\marginparwidth}{\raggedright\normalfont\scriptsize\url{#1}}}}
\def\eq#1{(\ref{#1})}
\newcommand{\R}{\textcolor{red}}


\hypersetup{pageanchor=false}
			
\vspace{.4cm}
\begin{center}
\noindent{\Large \bf{Exact lattice bosonization of the double scaled $c=1$ matrix model}}\\
\vspace{1cm}
Adwait Gaikwad\footnote{adwait.gaikwad@gmail.com} $^{a,b}$,
Gautam Mandal\footnote{mandal@theory.tifr.res.in, gautam.mandal@icts.res.in} $^b$, and 
Ajay Mohan\footnote{ajay.mohan@tifr.res.in} $^{b,c}$
\vspace{.3cm}
\begin{center}
$^a$ {\it Mandelstam Institute for Theoretical Physics, School of Physics}\\
{\it University of the Witwatersrand, Johannesburg 2050, South Africa.}\\
{~}\\
$^b$ {\it International Centre for Theoretical Sciences} \\
{\it Shivakote, Bangalore 560089, India.}\\
{~}\\
$^c$ {\it Department of Theoretical Physics}\\
{\it Tata Institute of Fundamental Research, Mumbai 400005, India.}\\
\end{center}
\end{center}				
\gap2
\begin{abstract}
  In a recent paper \cite{Mandal:2024xqn}, henceforth called paper I, it has been shown that entanglement entropy (EE) of matrix quantum mechanics (or, equivalently, of a non-relativistic fermion model) has a natural interpretation in terms of an exact bosonization of the fermion theory in terms of a lattice boson with lattice spacing $\sim 1/N$ (where $N$ is the rank of the matrix, equal to the number of fermions). The theory of this lattice boson is distinct from a latticized version of the ``collective variable'' theory which is formulated in terms of the fermion density $\rho(\lambda)$  (``collective variable''); it is, in fact, more closely related to the Wigner distribution of the fermions. In this paper, we develop this connection further for the $c=1$ double scaled matrix model. Even though $N$ becomes infinite, the double scaled theory is described by a boson on a finite and non-uniform lattice spacing, of order $\sim g_s(x) l_s$ (where $g_s(x) \equiv g_s {\rm cosech}^2(x/l_s)$ is a position dependent string coupling). We explicitly derive the $c=1$ EE, as well as the one and two-point functions of fermion density from the lattice boson. The results are explicitly finite, unlike the collective variable theory which is shown to yield divergent answers. Our results suggest that while local observables such as the EE and $\lan \rho(\lambda) \ran$, are characterized by a unique short distance scale $g_s(x) l_s$, the asymptotic observables such as the S-matrix at every genus are characterized by a short distance scale $l_s$. We show how this arises and discuss a possible interpretation in terms of D0 brane physics.
  
\end{abstract}

\pagenumbering{roman}

\newpage

\tableofcontents
\pagenumbering{arabic}
\setcounter{page}{1}



\section{Introduction and Summary}\label{sec:intro}

Two-dimensional string theory, in spite of the low dimensionality, continues to provide fertile ground for exploring ideas related to nonperturbative gravity. As is well-known, this theory is dual to the c=1 matrix model, which is a theory of $N \times N$ time-dependent hermitian matrices (for reviews of the original duality, obtained from the viewpoint of dynamically triangulated random surfaces, see, e.g., \cite{Klebanov:1991qa, Das:1992dm, Ginsparg:1993is, Jevicki:1993qn, Polchinski:1994mb}; this correspondence has been reinterpreted in the language of unstable D0 branes in \cite{McGreevy:2003kb, Martinec:2003ka, Klebanov:2003km, Douglas:2003up, McGreevy:2003dn, Takayanagi:2003sm}; recently nonperturbative evidence of this duality has appeared in \cite{Balthazar:2016utu, Balthazar:2017mxh,  Balthazar:2018qdv, Sen:2019qqg, Balthazar:2019ypi, Balthazar:2019rnh}). The matrix model is considered to be in the singlet sector (in the D0-brane version, this is automatic since the non-singlet sector is gauged away), and becomes a theory of $N$ non-relativistic fermions moving in an external potential.\footnote{For recent studies of the non-singlet sector of the matrix quantum mechanics \cite{Cho:2024kxn}, and references therein.} 

An important tool to describe the above duality has been the collective variable description of the fermion theory \cite{Das:1990kaa, Sengupta:1990bt}\footnote{``Collective variable'' refers to the fermion density or equivalently the eigenvalue density of the matrix.}, which correctly captures low energy fluctuations near the Fermi surface\footnote{These fluctuations in a metal are related to sound modes.}, and is described by an action of the form \eq{collective-action-0}. In this description, fluctuations of the collective variable are identified with the massless scalar field of the 2D string theory (famously misnamed the tachyon).\footnote{Up to a non-local wavefunction renormalization called a ``leg-pole factor''.} This identification is supported by the agreement at tree-level between amplitudes of the collective variable and those of the tachyon. This, in a sense, is a concrete realization of the emergence of string fields from matrices which can be regarded as a precursor of AdS/CFT.

Unfortunately because of the inherent limitation of the collective variable description (namely that it is supposed to describe low energy fluctuations near the Fermi surface), quantum entanglement in the ground state as well as loop amplitudes are not captured faithfully by the collective variable theory. In particular, both these quantities are divergent in the collective theory whereas they are finite in the fermion theory (the finite loop amplitudes are reproduced by string theory; the entanglement entropy (EE) is not yet computed from string theory).

It turns out that the quantities mentioned above (loop amplitudes and EE) involve in a nontrivial way the nonrelativistic nature of the fermion theory. Thus, a bosonic description that can faithfully reproduce these quantities, must involve a bosonization of nonrelativistic fermions. The first step towards this was achieved in \cite{Dhar:2005fg} where a set of $N$ bosonic oscillators $a_k, a^\dagger_l$ was introduced such that all operators in an $N$-fermion Hilbert space could be mapped in a one-to-one fashion to these bosonic oscillators (see Appendix \ref{app:op-boson} for more details). In \cite{Mandal:2024xqn}, henceforth called paper I, we constructed a bosonic field on a lattice (see \eq{real-space-boson}) corresponding to these oscillators. The question now boils down to whether this new bosonic formulation correctly reproduces the finite loop amplitudes and EE of the fermion theory. It is worth mentioning here that this bosonic formulation \eq{real-space-boson} is related to the earlier non-abelian bosonization of the fermions in terms of the Wigner distribution \cite{Dhar:1992hr, Dhar:1992rs, Das:1991ba, Das:1991uta}; details of this correspondence are presented in Appendix \ref{app:lattice-vs-uxp}.

In paper I, we showed in detail that the lattice boson has an in-built cut-off with a lattice spacing $\sim 1/N$, leading to a finite EE as well as a finite one-loop expectation value of the eigenvalue density $\rho(\l)$ (same as the fermion density $\psi^\dagger(\l) \psi(\l)$). We showed that both these quantities matched with those of the fermion theory. 

In this paper, we will focus on the $c=1$ matrix model. It was already noted in paper I that  although $N\to\infty$ in this case, the lattice spacing does not vanish owing to double scaling. It was also noted in paper I that if one computes the finite EE {\it prior to double scaling} and takes the double scaled limit of that expression, the limit remains finite and matches the Fermion result. In this paper, we will derive the double scaled theory in detail, and show that the loop corrections to one- and two-point functions of $\rho(\l)$ as well as the EE, computed in this theory, turn out to be finite (owing to the finite lattice spacing) and equal to the results from the fermionic theory (or, equivalently the martix model). In our discussion of the EE, we present subleading terms which appear to have the correct dependence on string coupling as expected from higher genus corrections. We will also present for comparison computation of these quantities from collective variable theory which turn out to be divergent.

A section-wise summary of the rest of the paper is as follows. In Section \ref{sec:c=1-boson} we present in detail the lattice boson formulation of the double scaled $c=1$ theory. We show that in the double scaling limit, the normal mode expansion reflects an infinite, but discrete lattice in coordinate space (corresponding to continuous ``momenta'' with a UV cut-off, see \eq{normal-2}, \eq{normal-3}) with non-uniform lattice spacing (see \eq{normal-2}, \eq{normal-3}). In Section \ref{sec:ee-and-subleading}, we present the EE calculation from the lattice boson theory and show that the leading result \eq{leading-ee} agrees with the well-known reading result from the fermion theory. In Section \ref{sec:ee-fermion} we present subleading corrections to the EE in the double-scaled theory; the corrections are consistent with string theory genus expansion where the coupling constant $g_s(x)$ is position dependent. In Section \ref{sec:rho-vev} we first present the vacuum expectation value of the fermion density $\lan \rho(\l) \ran$ from the lattice, arriving at the formula \eq{1pt-lattice} which coincides with the formula obtained from the fermion theory. In Section \ref{sec:rho-fermi} we present the known leading result for the one-point function \eq{Moore-rho}; we then go on to derive corrections \eq{subleading-1pt} to the one-point function coming from the leading WKB corrections to the wavefunctions. The leading correction has the form of a string theoretic genus expansion in terms of a position dependent coupling constant $g_s(x)$ like the corrections to the EE. Importantly, the corrections are finite. In Section \ref{sec:rho-coll}, we present a computation of quantum correction to the one-point function in collective field theory. Because of the cubic interactions, the quantum correction involves a tadpole diagram, which gives a UV-divergent computation; this shows limitations of the collective field theory. In Section \ref{sec:rho-rho} we present the two-point function of the density field. As before, we first compute it from the lattice-boson theory. The final result \eq{rho-rho-c} agrees with the fermionic result. Using this result, one can derive the on-shell 1-loop result \eq{xi-yin-2pt} for the two-point function \cite{Balthazar:2017mxh}. The result is finite and matches exactly the torus computation of the same quantity. An important difference from the EE and $\lan \rho(\l)\ran$, which were both local quantities and depend on the local genus expansion parameter $g_s(x)$, \eq{xi-yin-2pt} depends on the string coupling $g_s$ and $l_s$ separately. In Section \ref{sec:coll-2pt} we present a collective field theory computation of this quantity and find it to be UV-divergent; once again we find that collective field theory, in all these instances, does not compute quantum corrections correctly, whereas our lattice boson yields finite results which coincide with both the string theoretical computationa and the fermion theory computation. Finally we end with some concluding remarks.

\section{Double-scaled $c=1$ lattice boson}\label{sec:c=1-boson}

We begin this section with a brief review of our recent work on the exact {\em lattice bosonization} of matrix quantum mechanics \cite{Mandal:2024xqn}. It is a known fact that the class of time-dependent matrix models (also known as matrix quantum mechanics)
\begin{align}
S = \frac{\beta N}{l_s} \int dt \; \left\{\frac{l_s^2}{2}  \Tr \dot{M}^2 - \Tr V(M) \right\} \label{eqn:matrix-model-action}
\end{align}
is equivalent to the system of $N$ non-interacting, non-relativistic fermions living in the same potential as the matrix model (the eigenvalues of $M$ acting like fermions). The single particle Hamiltonian of this fermionic system is given by
\begin{align}
h = \frac{1}{l_s} \left(-\frac{1}{2}\frac{1}{(\beta N)^2} \frac{\partial^2}{\partial \lambda^2} + V(\lambda)\right) = \frac{1}{l_s} \left( \frac{p^2}{(\beta N)^2} + V(\lambda) \right)
\end{align}
with $(\beta N)^{-1}$ playing the role of $\hbar$. The $l_s$ factors in the matrix model action \eqref{eqn:matrix-model-action} are carefully accounted for in Appendix \ref{app:ls-factor}.

For such systems, we prescribe\footnote{This prescription is a natural extension of an earlier work on exact bosonization of non-relativistic fermions in \cite{Dhar:2005fg} and we would like to stress that this prescription is different from the lattice boson proposed in that paper.} a lattice boson constructed out of just $N$ bosonic oscillators $\{a_k\}$. This bosonic field lives on a lattice of $N$ points (labelled by $x_j$) and the normal mode expansion looks like\footnote{Compared with the lattice boson formulation in paper I, we have rescaled $\phi$, $\pi$, $H$ (the lattice boson Hamiltonian) and $t$ to absorb the trivial $\omega_0$ dependence that previously appeared in paper I.}
\begin{align}
\phi_j(t)  &\equiv \phi(x_j, t) = \sum_{n=1}^N \frac{1}{\sqrt{ \omega_n}} \sin \left(\frac{n \pi f(x_j)}{L} \right) [ a_n e^{-i \omega_n t} + a_n^{\dagger} e^{i \omega_n t} ]  \\
\pi_j(t) &\equiv \pi(x_j, t) = \sum_{n=1}^N \frac{i}{N}  \sqrt{\omega_n}  \sin \left(\frac{n \pi f(x_j)}{L} \right) [ -a_n e^{-i \omega_n t} + a_n^{\dagger} e^{i \omega_n t} ]
\label{real-space-boson} \\
&  \text{where } \quad \omega_n = \frac{2(N+1)}{\pi} \sin\left( \frac{n \pi}{2(N+1)} \right)
\end{align}
and $f(x)$ (or technically the inverse of it) encodes the non-uniform lattice structure. It was found that the exact boson corresponding to free fermions in zero potential (with hard box boundary conditions) lives on a uniform lattice (denoted by $y$) and that if it were any other varying potential, it would break the homogeneity of space and consequently the bosons would live on a non-uniform lattice (denoted by $x$). To find this non-uniform lattice, we first take the unique canonical ($W_{\infty}$) transformation that generates the change in potential from hard-box (of length $L$) to the required $V(\lambda)$ and we then project it on to the Fermi surface to get
\begin{align} \label{eqn:gen-winfty-trans}
y = f(x) \equiv \frac{L}{N \pi \hbar} \int_{x_-}^x dx' \; p^{}_f(x'), \; \; \; \;  \;\;\;\;\;  x_j = f^{-1}(y_j = j \epsilon)
\end{align}
where $x_j$'s mark the $N$ lattice points of this non-uniform lattice and $p^{}_f(x) = \sqrt{2(E_F - V(x))}$ is the Fermi momentum.

Let us now apply this framework to the $c=1$ model. After the double-scaling limit, the system becomes equivalent to fermions sitting on one side of an inverted harmonic oscillator potential with the Fermi sea at a depth $\mu$ below the potential maxima.
In other words, the Fermi energy is $-\mu$ and the Fermi momentum is $p_{f} = \sqrt{-2 \mu l_s + \lambda^2}$.
Based on previous work on the correspondence between the fluctuations of the Fermi surface of the double-scaled $c=1$ model and the tachyon field of the 2D string theory, a map from $\lambda$ to the string theory coordinate $x$ has already been established (along with the relation between $\mu$ and the string coupling $g_s$).
\begin{align}
  \lambda = \sqrt{2 \mu l_s} \cosh(x/l_s), \qquad \mu = \frac{1}{g_s l_s}
  \label{lam=cosh}
\end{align}
The lattice structure of our exact boson for the $c=1$ model in this $x$ coordinate is determined as follows
\begin{gather}
\frac{N \pi }{L} d y=\frac{ p_f(y) \;d y}{\hbar} = \frac{p_f(\lambda) \;d\lambda}{\hbar} = 2 \mu \sinh^2(x/l_s) \; dx, \qquad p_f(\lambda) = \sqrt{-2 \mu l_s + \lambda^2} \\
\Rightarrow y  = f(x) = \frac{2L\mu l_s}{N \pi}\left( \frac{\sinh(2x/l_s)}{4} -\frac{x}{2l_s} \right), \qquad y_j = j \frac{L}{N} = f(x_j) \\
\Rightarrow x_{j+1}-x_j = \frac{1}{2 \mu \sinh^2(x_j/l_s)} = \frac{g_s l_s}{2} \frac{1}{\sinh^2(x_j/l_s)} \equiv g_s(x_j)l_s =: \ep(x_j) \label{ep-def}
\end{gather}
The lattice is non-uniform and at large $x$ (by approximating $\sinh(x) \sim e^x$), the lattice points are given by
\begin{gather}
x_j \approx \frac{l_s}{2} \log(\frac{j}{\mu}) = \frac{l_s}{2} \log(j g_s) \quad \text{ and } \quad f(x_j) \approx \frac{L \mu l_s}{N \pi} e^{2 x_j/l_s}
\end{gather}
\begin{figure}[H]
\begin{center}
  \includegraphics[scale=0.4]{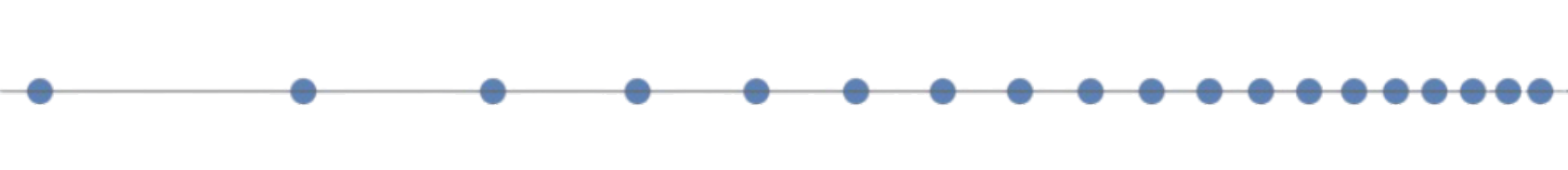}
  \caption{\small The non-uniform lattice of the double-scaled $c=1$ lattice boson.}
\end{center}
\end{figure}
\noindent This is a highly {\em non-trivial finding:}  despite setting $N = \infty$, our lattice boson continues to retain a (non-uniform) lattice structure\footnote{This is very special to the double-scaling limit of the $c=1$ model. For instance, this doesn't occur in ordinary potentials like the simple harmonic oscillator, where the lattice collapses to a continuum in the $N \to \infty$ limit.} with the effective lattice spacing given by $\ep(x)=g_s(x) l_s$ (see \eq{ep-def}).  With the lattice structure determined, we can now proceed to write the normal mode expansion of the bosonic fields.
\begin{align}
\phi(x_j) &= \sum_{n=1}^N \frac{1}{\sqrt{\omega_n}} \left( \sin(\frac{n \pi}{L} \frac{L \mu l_s}{N \pi} e^{2 x_j/l_s}) a_n  + \textit{h.c.} \right) \label{normal-1} \\
&= \int_0^N dn\; \frac{1}{\sqrt{\omega_n}} \left( \sin( \frac{n \mu l_s}{N} e^{2 x_j/l_s}) a_n + \textit{h.c.} \right) \nonumber \\
&= \int_0^{\mu} d k \; \frac{1}{\sqrt{\tilde \omega( k)}} \left( \sin( k l_s e^{2 x_j/l_s}) \tilde a ( k) + \textit{h.c.}  \right) \label{normal-2} \\
\pi(x_j) &= \sum_{n=1}^N \frac{i}{N} \sqrt{\omega_n} \left(- \sin\left( \frac{n \pi}{L} \frac{L \mu l_s}{N \pi} e^{2 x_j} \right) a_n  + h.c. \right) \\
&=\int_0^{\mu} d k \sqrt{\tilde \omega(k)}\left( - i \sin(k l_s e^{2 x_j}) \tilde a(k) + \textit{h.c.} \right) \label{normal-3}
\end{align}
where 
\begin{align}
  k &\equiv \frac{n\mu}{N}, \qquad  \tilde \omega( k) \equiv \frac{\mu \omega_n}{N} = \frac{2 \mu}{\pi} \sin(\frac{ k \pi}{2\mu}), \nonumber\\
  \tilde a (k) & \equiv \sqrt{\frac{N}{\mu}} a_n, \quad \tilde a^{\dagger}(k) \equiv \sqrt{\frac{N}{\mu}} a^{\dagger}_n, \quad [\tilde a(k),\tilde a^\dagger(k')]= \delta(k-k')
  \label{heisen}
\end{align}
We have used the Euler-Maclaurin formula to convert the sum into an integral and just as in the lattice structure, $N$ completely disappears from the normal mode expansion (in these new variables). The momentum modes are continuous but nevertheless come with a built-in cutoff $\mu$.

\subsection{The bosonic hamiltonian}\label{lattice-ham}
The lattice boson Hamiltonian, given in terms of the fermion single-particle eigenspectrum $\epsilon(m)$ ($m=0,1,\ldots$) as follows
\begin{align}
H = \sum_{j=1}^N \epsilon\left(\sum_{i=j}^N a_i a_i^{\dagger}\right)  
\end{align} 
becomes the following in the double-scaling limit
\begin{align}
  \frac{H}{N} = \frac{1}{\mu} \int_0^{\mu} dk \;\epsilon \left( \int_k^{\mu} d\tilde k \; \frac{1}{\mu}\tilde a(\tilde k) \tilde a^{\dagger}(\tilde k) \right)
  \label{lattice-ham}
\end{align} 

\section{Entanglement entropy and subleading corrections}\label{sec:ee-and-subleading}

In this section, we will discuss the ground state entanglement entropy (EE) of an interval $(x_1, x_2)$ in the double scaled $c=1$ theory. First, we will discuss how the leading EE \eq{leading-ee} (obtained originally from the fermion theory) comes out in the lattice boson theory. Next, we turn to the fermion theory; since the leading term has been analyzed before, we will describe computation of subleading corrections from the fermionic formulation.

\subsection{EE from lattice boson theory for the double scaled $c=1$ model}\label{sec:ee-lattice}

In our lattice boson theory, we compute entanglement entropy using the method described in \cite{Casini:2009sr}. It involves first computing the two-point correlators $X_{ij} = \braket{\phi(x_i) \phi(x_i)}$ and $Y_{ij} = \braket{\pi(x_i) \pi(x_j)}$ for those lattice points $x_j$ that are inside the entangling region $A$. Using (a) the normal mode expansions \eq{normal-2}, \eq{normal-3}, (b) the commutation relation \eq{heisen}, and (c) the fact that the ground state of the lattice hamiltonian \eq{lattice-ham} is $|0\ran$ defined by $a(k)|0 \ran=0$ \cite{Mandal:2024xqn, Dhar:2005fg}, we get
\begin{align}
\braket{\phi_i \phi_j} &= \int_0^{\mu} dk\; \frac{1}{\tilde \omega(k)} \sin(k e^{2 x_i}) \sin(k e^{2 x_j}) \\
\braket{\pi_i \pi_j} &= \int_0^{\mu} dk\; \tilde\omega(k)  \sin(k e^{2 x_i}) \sin(k e^{2 x_j}) 
\end{align}
where $\tilde \omega(k) \propto \mu \sin(k \pi/2\mu)$.
The EE is then given by the eigenvalues of the matrix $C = \sqrt{XY}$ as follows
\begin{align} \label{eqn:casini-huerta-ee}
S = \Tr \left[ \left( C + \frac{1}{2} \right) \log\left( C+\frac{1}{2}\right) - \left( C- \frac{1}{2} \right) \log\left( C-\frac{1}{2}\right) \right]
\end{align}
Importantly, these correlators are a function of only the labels of the lattice points. Therefore, one can compute them in the $y$ variable corresponding to the uniform lattice in \eqref{eqn:gen-winfty-trans} (associated to the fermions in a hard box potential). In that setting, the ground state EE has already been found to be
\begin{align}
S = \frac{1}{3} \log\left( \frac{N}{L} (y_2 - y_1) \right)
\end{align}
which can be transformed to the $x$ variable as
\begin{align}
  S &= \frac{1}{3} \log \left( \frac{x_2-x_1}{\ep(x_0)} \right),
  \label{leading-ee}
\end{align}
where
\begin{align}
  \ep(x_0) = \fr{g_s l_s}{\sinh^2\left(\frac{x_0}{l_s}\right)} \equiv g_s(x_0) l_s
  \label{short-dis}
\end{align}
defines the built-in short distance scale of the theory (cf. \eq{ep-def}). Eq. \eq{leading-ee} is a UV-finite result and it agrees with the leading EE from the fermion theory \cite{Das:1995vj, Hartnoll:2015fca}. Here $x_0$ is a point inside the interval $\{x_1,x_2\}$, $x_2-x_1=l$.\footnote{The precise location of $x_0$ does not matter since the interval is considered to be short enough so that the variation of $g_s(x)$ within the interval is small.}

\subsection{EE from the double scaled fermion theory}\label{sec:ee-fermion}

As mentioned above, the leading result for the entanglement entropy of the double-scaled $c=1$ model agrees between the lattice boson computation and the fermionic computation. The leading fermionic EE corresponds only to the second cumulant in the cumulant expansion. In the following, we explore the subleading corrections coming from the higher cumulants. 

In a simpler system, of fermions on a circle, these subleading corrections have already been identified \cite{Calabrese:2011zzb}
\begin{align}
  S = \frac{1}{3}\log(2 N \sin(\frac{\pi l}{L})) + \mathcal{O}\left(\frac{1}{N^2}\right)
  \label{subleading-ee}
\end{align}
which are of order $1/N^2$. We will analyse the form of these corrections below.

\paragraph{Scaling argument:} Note that all cumulants $V^{(k)}$ in the EE cumulant expansion 
\begin{align}
S = \frac{\pi^2}{3} V^{(2)} + \frac{\pi^2}{45} V^{(4)} + \cdots
\end{align}
can be written in terms of traces of the ``overlap matrix'' $A$  as follows \cite{Calabrese:2011ycz}:
\begin{gather}
A_{m n} =  \int_A dx\; \phi_m^*(x) \phi_n(x), \qquad m,n \in \text{occupied levels} \\
V^{(2)} = \Tr A - \Tr A^2, \hspace{7ex} V^{(4)} = \Tr A- 7 \Tr A^2 + 12 \Tr A^3 - 6\Tr A^4, \hspace{7ex} \text{etc.}
\end{gather}
Note also that all such traces can be expressed as integrals of the fermion bilinear 
\begin{align}
\Tr A^{n} &= \int_A dx_1 \cdots \int_A dx_n \; \bra{\Omega}\Psi^{\dagger}(x_1)\Psi(x_2)\ket{\Omega}  \cdots \bra{\Omega}\Psi^{\dagger}(x_n)\Psi(x_1)\ket{\Omega} \\
&= \int_0^l dx_1 \cdots \int_0^l dx_n \; \frac{1}{L}\frac{\sin( N \pi (x_1 - x_2)/L)}{\sin(\pi (x_1-x_2)/L)}  \cdots \frac{1}{L}\frac{\sin(N \pi (x_n - x_1)/L)}{\sin(\pi (x_n-x_1)/L)} \label{eqn:ferm-circle-trAn}
\end{align}
This is the ground state bilinear you obtain using the exact wavefunctions of free fermions on a circle: $\chi_n(x) = \exp(i 2\pi n \hbar x/L) / \sqrt{L}$. We will now work in the $l/L \ll 1$ limit which allows us to approximate the $\sin(\pi (x_{j-1} - x_j)/L)$ factors with just the argument $\pi (x_{j-1} - x_j)/L$ and thereby, enables us to rescale the integration variables to $x' = x/l$. 
\begin{align} \Tr A^n &= \int_0^l dx_l \cdots \int_0^l dx_n \; \frac{1}{\pi (x_1-x_2)} \sin(\frac{N \pi  (x_1-x_2)}{L} ) \cdots \frac{1}{\pi (x_n-x_1)} \sin(\frac{N \pi  (x_n - x_1)}{L}) \nonumber \\
&= \int_0^1 dx'_1 \cdots \int_0^1 dx'_n \; \frac{1}{\pi (x'_1-x'_2)} \sin(\frac{N \pi l (x'_1-x'_2)}{L} ) \cdots \frac{1}{\pi (x'_n-x'_1)} \sin(\frac{N \pi l (x'_n - x'_1)}{L}) \label{eqn:free-ferm-tr}
\end{align}
Expressed in this form, it is clear that entanglement entropy must be purely a function of $Nl/L = p_F l/\pi$ and that the order $1/N^2$ correction we saw earlier must in fact be $\mathcal{O}\left((Nl/L\right)^{-2})$. This scaling argument can be generalized to the case of fermions in a general potential where the bilinear in the semiclassical (Thomas-Fermi) limit takes the form
\begin{align}\label{eqn:largen-bilinear}
\bra{\Omega} \Psi^{\dagger}(x_1) \Psi(x_2) \ket{\Omega} = \frac{1}{\pi (x_2 - x_1)} \sin\left[ \frac{(x_2 - x_1)}{\hbar} p_F\left(\frac{x_1+x_2}{2} \right) \right]
\end{align}
provided our entangling region is small compared to the scale of variation of the potential so that we can replace the function $p_F$ with just $p_F(x_0)$ where $x_0$ is some point inside the entangling region.
One can now rerun the same argument as before (with $N \pi/L$ replaced by $p_F(x_0)$) to assert that entanglement entropy must purely be a function of $p_F(x_0) l$. Therefore, the subleading correction to EE must be given by  
\begin{align}
S_{\text{subleading}} \sim \frac{1}{(p_f(x_0)\, l)^2}.
\end{align}


More quantitatively, the EE of free fermions on a circle of length $L$, including the leading correction and discarding unimportant additive numerical constants, is given by\footnote{We obtained this result from modifying the EE result provided in \cite{Calabrese:2010mtj} for the XX model on an infinite one-dimensional chain. This problem can be mapped to fermions on a circle by
\begin{align}
l^{\rm XX-chain} \leftrightarrow N \qquad \qquad {k_F}^{\rm XX-chain} \leftrightarrow \frac{\pi l}{L}
\end{align}}
\begin{align}
\scalebox{0.9}{$\displaystyle S = \frac{1}{3} \log( N \sin(\frac{\pi l}{L}) ) - \frac{1}{12 N^2} \left( \frac{1}{5} + \cot^2\left(\frac{\pi l}{L}\right) \right) \xrightarrow[\text{small $l/L$}]{\text{large $N$}} \log(\frac{ N \pi l}{L}) - \frac{1}{12}\left( \frac{N \pi l}{L} \right)^{-2} + \cdots$}
\end{align}
Therefore, for the double-scaled $c=1$ fermions, we get
\begin{align}
S 
&\approx \frac{1}{3} \log( \frac{l}{g_s l_s} \sinh^2(x_0) ) - \frac{1}{12}\frac{g_s^2 l_s^2}{\sinh^4(x_0)} \frac{1}{l^2}
= \frac{1}{3} \log( \frac{l}{\ep(x_0)} ) - \frac{1}{12} \fr{\ep(x_0)^2}{l^2} 
\label{ee-fermi-expansion}
\end{align}
Note that the leading as well as subleading expressions for the EE (given by \eq{leading-ee}, \eq{ee-fermi-expansion}) are given in terms of the dimensionless ratio $\fr{l}{\ep(x_0)}$; thus, the result is characterized by a single length scale $\ep(x_0)$ (defined in \eq{short-dis}).

\section{One-point function of the eigenvalue density}\label{sec:rho-vev}

\subsection{Lattice boson computation}\label{sec:rho-lattice}
The eigenvalue density $\rho(\lambda)$ in our lattice boson theory is given by
\begin{align}
\rho(\lambda) = \sum_{m,n=0}^{\infty} \hat{F}_{mn} \chi^*_m(\lambda) \chi_n(\lambda) \label{eqn:rho-latt-boson}
\end{align}
where $\hat{F}_{mn}$ is the RHS of equation \eqref{exact-op-map}, i.e. $\psi^{\dagger}_m \psi_n$ written in terms of the bosonic oscillators. Since the Fermi ground state is mapped to the no-particle state $\ket{0}$ in the bosonic theory, we are interested in finding the expectation value $\bra{0} \rho(\lambda) \ket{0}$. Although the general $\hat{F}_{mn}$ is complicated, we will write below the expressions for some special cases (applicable only when acted on $\ket{0}$)
\begin{align}
\hat F_{n-2,n} \ket{0} &= \left\{\delta(-n+N+1){\sigma_1}^2 - \delta(-n+N) \sigma_2\right\}\ket{0} = 0 \label{eqn:latt-rho-1}\\
\hat{F}_{n-1 , n} \ket{0} &= \delta(-n+N)\sigma_1\ket{0} = 0  \label{eqn:latt-rho-2}\\ 
\hat F_{n,n} \ket{0} &= \sum_{j=1}^N \delta(N-n-j) \ket{0} \label{eqn:latt-rho-3}\\
\hat F_{n+1,n} \ket{0} &= \delta(-n + N -1) \sigma_1^{\dagger}\ket{0} = \delta(-n+N-1) \ket{r_1 = 1} \label{eqn:latt-rho-4} \\
\hat F_{n+2,n} \ket{0} &= \left\{\delta(-n+N-1) {\sigma_1^{\dagger}}^2 - \delta(-n+N-2) \sigma_2^{\dagger}\right\}\ket{0} \nonumber \\
& \qquad \qquad \qquad   = \delta(-n+N-1) \ket{r_1=2} - \delta(-n+N-2) \ket{r_2=1} \label{eqn:latt-rho-5}
\end{align}
where $\sigma_i$ is defined in Appendix \ref{app:op-boson}.
It is clear that only $\hat F_{n,n}$ return the state $\ket{0}$ and therefore, 
\begin{align}
\braket{\rho(\lambda)} &= \sum_{n=0}^{\infty} \braket{ \hat F_{n,n} } |\chi_n(\lambda)|^2 = \sum_{n=0}^{\infty} \sum_{j=1}^{N} \delta(-n +N-j)|\chi_n(\lambda)|^2 =  \sum_{j=0}^{N-1} |\chi_n(\lambda)|^2
\end{align}
where the final sum is over the occupied levels.
We can now perform the double-scaling limit to write in terms of the continuum energy variable $\nu$
\begin{align}
  \braket{\rho(\lambda)} = \int d\nu \; \Theta(\nu-\mu ) |\chi_{\nu}(\lambda)|^2  = \int_{\mu}^{\infty} d\nu \; |\chi_{\nu}(\lambda)|^2
  \label{1pt-lattice}
\end{align}
which agrees with the expression from the fermionic theory \cite{Moore:1991sf}. Interestingly, although the integration range is infinite, the result of the integral is still finite because the $\infty$ limit of the integral (for any fixed $\lambda$) corresponds to moving deeper and deeper into the forbidden region of the potential, where the wavefunction is exponentially damped. 

\subsection{From the double-scaled fermion theory}\label{sec:rho-fermi}

As we noted above, the expression \eq{1pt-lattice} for the ground state expectation value of $\rho(\l)$ matches between the lattice boson expression obtained above and that from the double scaled fermion theory. The fermionic computation was performed in \cite{Moore:1991sf} using the leading WKB approximation for the wavefunctions and the result, in two asymptotic regimes, was 
\begin{align}
\braket{\rho(\lambda)} = \begin{cases} \frac{1}{2 \pi} \sqrt{\lambda^2 - 4 \mu} = \fr1{\sqrt{g_s(x)}}  & \lambda^2 \gg 4 \mu \\ \frac{1}{2 \pi (\pi \mu)^{1/4}}\exp{- \pi \mu + 2 \sqrt{\mu} \lambda} & \lambda^2 \ll \mu \end{cases} \label{Moore-rho}
\end{align}
Note that the weak-coupling result on the top line is of the form
\begin{align}
  \braket{\rho(x)} =\braket{\rho(\lambda)} \frac{d \lambda}{dx} = \frac{1}{g_s(x) l_s} = \fr{1}{\ep(x)}
  \label{rho-ep}
\end{align}

It is not difficult to compute subleading WKB corrections to the wavefunctions. We present below the expression for the derivative of the density, $\lan\del\rho(\l)/\del\mu \ran$, including the first subleading correction: 
\begin{align}
-\frac{\partial \braket{\rho(\lambda)}}{\partial \mu}= |\psi_{\mu}(\lambda)|^2 &=  \frac{1}{\sqrt{\lambda^2 - 2 \mu}} \left( 1 - \frac{1}{64 l_s^2} \frac{\lambda^2 + 8 \mu}{(\lambda^2-2\mu)^3} + \cdots \right) \nonumber \\
&= \sqrt{\epsilon(x)} \left(1 - \frac{\epsilon(x)^2}{64 l_s^2} (1 + \mathcal{O}(e^{-x/l_s})) + \cdots \right)
\label{subleading-1pt}
\end{align}
where 
$\epsilon(x)$ is the lattice spacing of our lattice boson. Note that the correction has the form of an expansion in $\epsilon(x)^2$, as in the case of the entanglement entropy \eq{ee-fermi-expansion}, which is another local quantity. By contrast, as we will discuss in the next section, asymptotic observables like the S-matrix depend separately on $g_s$ and $l_s$ rather than on the combined factor $\epsilon(x) = g_s(x) l_s$.


\subsection{From collective field theory}\label{sec:rho-coll}

Note that the expectation value $\lan \rho(\l)\ran$ computed above is finite. We will now compare this calculation with what we get from collective field theory.

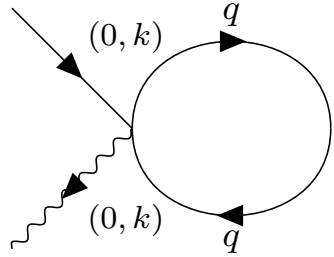
\begin{figure}[H]
\centering
\scalebox{1.5}{
\begin{tikzpicture}
  \begin{feynman}
    \vertex (a) ;
    \vertex [below right=of a] (b);
    \vertex [below left=of b] (f1) ;
    \vertex [right=1.75cm of b] (c);

    \diagram* {
      (a) -- [fermion, edge label=\footnotesize {\scriptsize \(  (0,k) \)}] (b) -- [charged boson, edge label = {\scriptsize \( (0,k)\)}] (f1),
      (b) -- [fermion, half left, edge label = \footnotesize \( q \)] (c) -- [fermion, half left, edge label = \footnotesize \( q \)] (b),
    };
  \end{feynman}
\end{tikzpicture}}
  \caption{\small Tadpole diagram. The squiggly line denotes momentum insertion from the interaction vertex $g_s(x)$.}
\end{figure}
\noindent The collective field theory action  (in the $\Pi_{\varphi}$, $\varphi$ language) is given by \eqref{coll-action1-app}
\begin{align}
S = \int dt \,d\lambda\, \left[ \Pi_\varphi \dot\varphi - \frac{1}{2} \Pi_\varphi^2 \varphi' - \frac{\pi^2}{6} \varphi'^3 + V(\l) \varphi' \right], \label{collective-action-0}
\end{align}
which, after eliminating $\Pi_{\varphi}$ using its equation of motion and performing a background subtraction $\varphi = \varphi_0 + \phi$,\footnote{See e.g. \eqref{fluct} and \eqref{coll-action5-app} in \autoref{app:collective-action-derivation} for more details.} becomes (working in $l_s =1$ units)\eqref{coll-action5-app}
\begin{align}
\scalebox{0.9}{$\displaystyle S = 2\pi \int dt\, dx \, \left\{ \partial_+ \phi \partial_- \phi - \frac{\pi g_s}{6 \sinh^2 x} \left( (\partial_+ \phi)^3 - (\partial_-\phi)^3\right) + \frac{\pi^2 g_s^2}{4 \sinh^4x} \left( (\partial_+ \phi)^2 - (\partial_-\phi)^2 \right)^2 \right\} + \mathcal{O}(g_s^3)$}
\label{collective-action}
\end{align}
where the first term is simply the free scalar field theory action written in $z_{\pm} = t \pm x$ coordinates and the second and the third terms are collectively the interaction terms $S_{\rm int}$ up to second order in $g_s$. The eigenvalue density $\rho$ is to be identified with $\partial_{\lambda} \varphi$. The derivation of this action is provided in Appendix \ref{app:collective-action-derivation}. Applying perturbation theory to first order in $g_s$, we get
\begin{align}
\braket{\partial_x \phi} &= i \braket{\partial_x \phi S_{\rm int}}_0 + \mathcal{O}(g_s^2) \\
&= - i g_s \frac{\pi}{2}  \int dt' \,dx'\, \frac{1}{\sinh^2 x'} \braket{\partial_x \phi \partial_+ \phi(t',x')} \braket{\partial_+ \phi(t',x') \partial_+ \phi(t',x')}
\label{real-quad}
\end{align}
If we call $g_s(x) = g_s/\sinh^2 x$, the external leg of the tadpole diagram $\braket{\partial_x \phi(t,x) \partial_+ \phi(t',x')}$ contributes a $g_s(x)$ factor that is localized at the point $x$ where the one-point function is evaluated.
\begin{align}
\int dt' dx'\; g_s(t',x') \braket{\partial_x \phi(t,x) \partial_+ \phi(t',x')} &= \int dk\; \frac{k^2}{k^2 - i \epsilon} \tilde g_s(k) e^{i k x} \\
&=\int dk \; \tilde g_s(k) e^{ikx} = g_s(x)
\end{align}
The remaining part is the usual tadpole divergence arising from the coincident propagator.
\begin{gather}
\braket{\phi(z)\phi(z')} = \log((z_+ - z_+')(z_--z_-'))
\implies \braket{\partial_+ \phi(z) \partial_+ \phi(z+\epsilon)} \sim \frac{1}{\epsilon_+^2} 
\end{gather}
This can also be seen in the following momentum integral corresponding to $\braket{\partial_+ \phi(x') \partial_+ \phi(x')}$
\begin{align}
I &= \int d^2q \frac{q_+^2}{q^2 - i \epsilon} \\
&= \frac{1}{2} \int dq_+ q_+ \int dq_- \frac{1}{q_- - \frac{i \epsilon}{q_+}} \\
& = \frac{1}{2} \int dq_+ q_+ \int dq_- \left[ P\left(\frac{1}{q_-}\right) + i \pi \delta(q_-) {\rm sgn}(q_+) \right] \\
&= \frac{i \pi}{2} \int dq_+ q_+ {\rm sgn}(q_+) \to \frac{i \pi}{4} \int_0^{\Lambda_+} dq_+ q_+ \sim \Lambda_+^2
\label{mom-quad}
\end{align}
Put together, 
\begin{align}
\braket{\partial_x \phi} = c \,g_s(x) \Lambda_+^2 = c \, \frac{\Lambda_+^2}{\rho_0(\lambda)^2} 
\end{align}
$c$ is a numerical factor and the degree of divergence is the same as one expects from power counting. In the context of the two-point function, we will find in the next section that because of the light-cone form of the vertices, the divergence can be milder than what naive power counting suggests. The result of $\braket{\partial \rho/\partial \mu}$ from collective field theory
\begin{align}
-\frac{\partial \braket{\rho(\lambda)}}{\partial \mu} &= -\frac{\partial \rho_0}{\partial \mu} - \frac{\partial}{\partial \mu} \left( \frac{1}{\rho_0(\lambda)} \partial_x \phi \right) \\
&= \frac{1}{\rho_0(\lambda)} - \frac{3 c}{\rho_0(\lambda)^5} \Lambda_+^2 = \sqrt{\epsilon(x)} \left( 1 - 3 c \,\epsilon(x)^2 \Lambda_+^2 \right)
\end{align}
has the same parametric form in $\epsilon(x)$ as the fermion theory result \eqref{subleading-1pt}, except that the $\Lambda_+^2$ divergence is replaced by a finite contribution in the fermion theory calculation (as one would expect from a string theory computation).


%

\section{Two-point function of the eigenvalue density}\label{sec:rho-rho}

\subsection{Lattice boson computation}\label{sec:lattice-2pt}

Just like in the one-point function case, we are now interested in computing 
\begin{align}
\bra{0} \rho(\lambda_1) \rho(\lambda_2) \ket{0} = \sum_{p,q,m,n=0}^{\infty} \bra{0} \hat F_{p q} \hat F_{m n} \ket{0} \chi^*_p(\lambda_1) \chi_q(\lambda_1) \chi^*_m(\lambda_2) \chi_n(\lambda_2) 
\end{align}
where we have used the lattice boson definition of $\rho(\lambda)$ in \eqref{eqn:rho-latt-boson}. Looking at the special cases of the application of $\hat F_{m n}$ on $\ket{0}$ in equations \eqref{eqn:latt-rho-1} to \eqref{eqn:latt-rho-5}, we see that $\hat F_{mn} \ket{0} = 0$ for $m < n$ which implies that only $m \geq n$ contributes to the double sum. Furthermore, we see that only the following combinations contribute
\begin{align}
\bra{0} \hat F_{q-j,q} \hat F_{n+j,n} \ket{0}, \qquad j \geq 0
\end{align}
For example, if we have $\hat F_{n+1,n} \ket{0} = \delta(-n + N-1) \,\sigma_1^{\dagger}\ket{0}$ from the right, then we need $\bra{0} F_{q-1,q} = \bra{0} \sigma_1\, \delta(-q+N) $ from the left to give a non-zero inner product. Hence we get contributions like
\begin{align}
&\sum_{q,n=0}^{\infty} \braket{ \hat F_{q,q} \hat F_{n,n}} |\chi_q(\lambda_1)|^2 |\chi_n(\lambda_2)|^2 = \sum_{q=0}^{N-1}  \sum_{n=0}^{N-1} |\chi_q(\lambda_1)|^2 |\chi_n(\lambda_2)|^2 \\
&\sum_{q,n=0}^{\infty} \braket{ \hat F_{q-1,q} \hat F_{n+1,n}} \chi^*_{q-1}(\lambda_1)\chi_q(\lambda_1) \chi_{n+1}(\lambda_2) \chi_n(\lambda_2) \nonumber \\
&\qquad \qquad \qquad \qquad= \sum_{q,n=0}^{\infty} \delta(-q+N) \delta(-n+N-1) \chi^*_{q-1}(\lambda_1) \chi_q(\lambda_1) \chi^*_{n+1}(\lambda_2) \chi_n(\lambda_2) \nonumber \\
 & \qquad \qquad \qquad \qquad = \chi^*_{N-1}(\lambda_1) \chi_N(\lambda_1) \chi^*_N(\lambda_2) \chi_{N-1}(\lambda_2)  \\ 
 &\sum_{q,n=0}^{\infty}\braket{\hat F_{q-2,q} \hat F_{n+2,n} } \chi^*_{q-2}(\lambda_1)\chi_q(\lambda_1) \chi^*_{n+2}(\lambda_2) \chi_n(\lambda_2) \nonumber \\
 &\qquad \quad = \begin{multlined}[t] \sum_{q,n=0}^{\infty} \bra{0}\{\sigma_1^2\delta(-q+N+1) -\sigma_2 \delta(-q+N)\} \\ \{ (\sigma_1^{\dagger})^2 \delta(-n+N-1) - \sigma_2^{\dagger} \delta(-n+N-2) \} \ket{0} \chi^*_{q-2}(\lambda_1)\chi_q(\lambda_1) \chi^*_{n+2}(\lambda_2) \chi_n(\lambda_2) \end{multlined} \nonumber\\
 &\qquad \quad =\chi^*_{N-1}(\lambda_1)\chi_{N+1}(\lambda_1) \chi^*_{N+1}(\lambda_2) \chi_{N-1}(\lambda_2) + \chi^*_{N-2}(\lambda_1)\chi_N(\lambda_1)\chi^*_N(\lambda_2) \chi_{N-2}(\lambda_2)
\end{align}
The pattern we observe is that 
\begin{align}
&\sum_{q,n=0}^{\infty} \braket{ \hat F_{q,q} \hat F_{n,n} } |\chi_q(\lambda_1)|^2 |\chi_n(\lambda_2)|^2 = \sum_{q \,\in \, {\rm occupied}} \;\sum_{n\, \in \, {\rm occupied}}  |\chi_q(\lambda_1)|^2 |\chi_n(\lambda_2)|^2 \label{eqn:rho-rho-unconn}\\
&\sum_{j=1}^{\infty}\sum_{q,n=0}^{\infty} \braket{ \hat F_{q-j,q} \hat F_{n+j,n} } \chi^*_{q-j}(\lambda_1) \chi_q(\lambda_1) \chi^*_{n+j}(\lambda_2) \chi_n(\lambda_2) \nonumber \\
&\qquad \qquad \qquad \qquad \qquad= \sum_{q \, \in\, {\rm unoccupied}}\;\sum_{n \, \in \, {\rm occupied}} \chi^*_n(\lambda_1)\chi_q(\lambda_1) \chi^*_q(\lambda_2) \chi_n(\lambda_2) \label{eqn:rho-rho-conn}
\end{align}
It is clear that \eqref{eqn:rho-rho-unconn} is $\braket{\rho(\lambda_1)} \braket{\rho(\lambda_2)}$ and \eqref{eqn:rho-rho-conn} is the connected density-density correlator $\braket{\rho(\lambda_1)\rho(\lambda_2)}_{\rm c}$. In summary,
\begin{align}
  \braket{\rho(\lambda_1)\rho(\lambda_2)}_{\rm c} = \sum_{q \, \in\, {\rm unoccupied}}\;\sum_{n \, \in \, {\rm occupied}} \chi^*_n(\lambda_1)\chi_q(\lambda_1) \chi^*_q(\lambda_2) \chi_n(\lambda_2)
  \label{rho-rho-c}
  \end{align}

\subsection{$1 \to 1$ scattering amplitude from the double-scaled fermion theory}\label{sec:fermi-2pt}

The expression \eq{rho-rho-c} obtained from the lattice boson theory reproduces precisely the two-point function of the fermion density (see, e.g. \cite{Das:2022mtb}). In the double-scaled $c=1$ theory, this result has been used to derive the $1 \to 1$ tachyon scattering amplitude \cite{Balthazar:2017mxh}. As in the previous sections, we are interested in quantum corrections: the correction to the one-loop $1 \to 1$ scattering amplitude is given by
\begin{align}
  S_{1 \to 1}(\omega)= \frac{g_s^2}{24}(i \omega^2 + 2 i \omega^4 - \omega^5).
  \label{xi-yin-2pt}
\end{align}
Here we have used $l_s=1$ units. The 1-loop correction is clearly finite. In \cite{Balthazar:2017mxh}, it has been shown that the result exactly matches string world-sheet computations on the torus. We now investigate whether collective field theory reproduces this result. 

\subsection{Collective field theory computation of $1 \to 1$ scattering amplitude}\label{sec:coll-2pt}

We see from the form of the collective theory action in \eqref{collective-action} that two types of Feynman diagrams contribute at the one-loop level (order $g_s^2$), i.e. cubic interaction and quartic interaction. The amputated connected part of the one-loop two-point function is then given by \(S^{\rm coll.}_{1 \to 1} = I_{\rm cubic} + I_{\rm quartic}\).

\subsubsection{Cubic interaction}
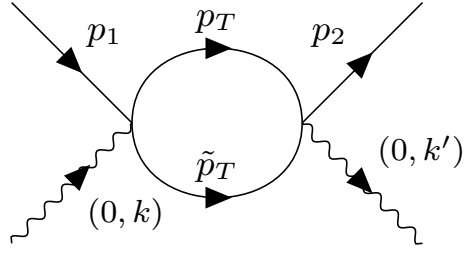
\begin{figure}[H]
\centering
\scalebox{1.5}{
\begin{tikzpicture}
  \begin{feynman}
    \vertex (a) ;
    \vertex [below right=of a] (b);
    \vertex [below left=of b] (f1) ;
    \vertex [right=of b] (c);
    \vertex [above right=of c] (f2);
    \vertex [below right=of c] (f3);

    \diagram* {
      (a) -- [fermion, edge label= \footnotesize \(  p_1 \)] (b) -- [anti charged boson, edge label = {\scriptsize \( (0,k)\)}] (f1),
      (b) -- [fermion, half left, edge label = \footnotesize \( p_T \)] (c),
      (b) -- [fermion, half right, edge label = \footnotesize \( \tilde p_T \)] (c),
      (c) -- [fermion, edge label = \footnotesize \( p_2 \)] (f2),
      (c) -- [charged boson, edge label = {\scriptsize \( (0,k') \)}] (f3),
    };
  \end{feynman}
\end{tikzpicture}}
\caption{\small Contribution to the two-point function from the cubic vertex. The squiggly lines denote the momentum coming from the position dependent coupling constant $g_s(x)$. As a result, $k_1 + k = k_2 + k'$, $ \omega_1 = \omega_2$, $p_T = (\omega_T, k_T)$ and $\tilde p_T= (\omega_1 - \omega_T, k_1 + k - k_T)$ where $p_i = (\omega_i, k_i)$. The on-shell conditions $\omega_i=k_i$ force $k= k'$.}
\end{figure}
\begin{equation}
\scalebox{0.92}{$\displaystyle I_{\rm cubic} = -C_1  \pi^6 \left(\f{1}{24\pi}\right)^2 p_1^+ p_2^+ \int dk d\omega_T dk_T\ \f{\tilde g_s(k)^2\ (p_T^+)^2(\tilde{p}_T^+)^2}{(\omega_T^2 - k_T^2 + i \ep)((\omega_1 - \omega_T)^2 - (k_1 + k - k_T)^2 + i \ep)}$} 
\end{equation}
We are computing the result up to an overall numerical factor $C_1$ and $\tilde g_s(k)$ is the Fourier transform of $g_s(x) = g_s/\sinh^2(x)$. 
\begin{align}
\tilde g_s(k) = -\frac{g_s}{\sqrt{2\pi}} \left(2 + k \pi \coth(\frac{k \pi}{2}) \right)
\end{align}
The above expression arises from the $(\partial_+ \phi)^3$ interaction term; the $(\partial_- \phi)^3$ interaction doesn't contribute under the on-shell conditions $\omega_1 = k_1 > 0$ and $\omega_2 = k_2 >0$. 
Redefining \(\omega = \omega_1 = \omega_2 = k_1 = k_2\) and \(p_T =p\), we get
\begin{align}
I_{\rm cubic} &= -\f{C_1 \pi^4}{288}\ \omega^2 \int dk dp^+\ \tilde g_s(k)^2\ (p^+)^2(2\omega + k - p^+)^2 \nonumber \\ 
&\qquad \qquad \qquad \qquad \qquad \int  dp^-\ \f{1}{(p^+p^- + i \ep)[(k-p^+)(2\omega - k - p^-)+i\ep]}   \\
&= i\f{C_1  \pi^5}{864}\ \omega^2 \int \f{dk}{k} \tilde g_s(k)^2\ (k + 2\omega)^3 
\end{align}
More details about this computation can be found in Appendix \ref{app:rho-rho}. The integration is UV divergent so we put a UV cutoff \(k \in (-\Lambda, \Lambda)\), which gives us 
\begin{equation}
\scalebox{0.9}{$\displaystyle I_{\rm cubic} = i\f{C_1 g_s^2 \pi^4}{540}\  \left[\omega^2 \left(\frac{\pi  \Lambda ^5}{8}+\frac{5 \Lambda ^4}{8}+\frac{5 \Lambda ^3}{6 \pi } + 1\right) + \omega^4 \left(\frac{5 \pi  \Lambda ^3}{2}+15 \Lambda ^2+\frac{30 \Lambda }{\pi }+20  \right) - i\ \omega^5\ 40   \right]$}
\end{equation}

\subsubsection{Quartic interaction}
\begin{figure}[!h]
\centering
\scalebox{1.5}{
\begin{tikzpicture}
  \begin{feynman}
    \vertex (a) ;
    \vertex [right=of a] (b);
    \vertex [below=of b] (f1) ;
    \vertex [right=of b] (c);
    \vertex [above=of b] (d);

    \diagram* {
      (a) -- [fermion, edge label= \footnotesize \( p \)] (b) -- [anti charged boson, edge label = {\scriptsize \( (0,\tilde k)\)}] (f1),
      (b) -- [fermion, half left, edge label = \footnotesize \( q \)] (d) -- [fermion, half left, edge label = \footnotesize \( q \)] (b),
      (b) -- [fermion, edge label = \footnotesize \( p \)] (c),
    };
  \end{feynman}
\end{tikzpicture}}
\caption{\small Contribution to the two-point function from the quartic vertex. The squiggly line denotes moomentum coming from the position dependent coupling constant $g_s(x)$. $p = (w,k)$ and the on-shell condition $\omega = k$ force $\tilde k =0$. }
\end{figure}
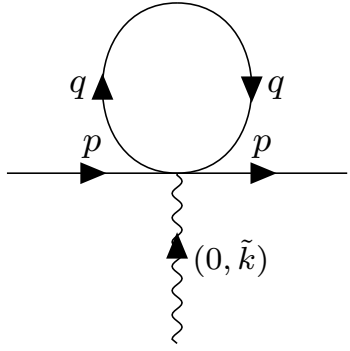
\noindent Only the interaction terms $(\partial_+ \phi)^4$ and $(\partial_+ \phi)^2 (\partial_-\phi)^2$ contribute under the on-shell condition $\omega = k >0$ and both of these quartic interactions give rise to a coincident propagator that is quadratically divergent
\begin{align}
I_{\rm quartic}  &=  C_2 \pi^6 \frac{\pi^2}{4} \frac{1}{4\pi^3} (p^+)^2   \widetilde{g_s^2}(0) \left[ \int d^2q\;    \frac{(q^+)^2}{q^2 - i \epsilon} - \frac{1}{3}  \int d^2q\;    \frac{(q^-)^2}{q^2 - i \epsilon}\right]  \\
&= C_2 \pi^5 \frac{4}{9 \sqrt{2 \pi}}  \omega^4 \Lambda^2 
\end{align}
where we have input the Fourier transform of $g_s^2(x)$
\begin{align}
\widetilde{g_s^2}(k) = \frac{g_s^2}{6\sqrt{2 \pi}} \left(8 + 6k^2 + 4 \pi k \coth(\frac{k \pi}{2}) + \pi k^3 \coth(\frac{k \pi}{2}) \right)
\end{align}
The cubic and quartic contributions taken together, we get
\begin{multline}
\scalebox{0.77}{$\displaystyle S^{\rm coll.}_{1\to 1}(\omega)=   i\f{C_1 g_s^2 \pi^4}{540}\  \left[\omega^2 \left(\frac{\pi  \Lambda ^5}{8}+\frac{5 \Lambda ^4}{8}+\frac{5 \Lambda ^3}{6 \pi } + 1\right) + \omega^4 \left(\frac{5 \pi  \Lambda ^3}{2}+15 \Lambda ^2+\frac{30 \Lambda }{\pi }+20  \right) - i\ \omega^5\ 40   \right] + C_2 \pi^5 \frac{4}{9 \sqrt{2 \pi}}  \omega^2 \Lambda^2 $}
\end{multline}


\section{Concluding remarks}\label{sec:conclusion}

In this section, we will elaborate a specific aspect of our results which we find to be of particular interest and are not summarized in Section \ref{sec:intro}.\footnote{We are grateful to Ashoke Sen for several illuminating discussions on this issue.}

We found in this paper that our results \eq{ee-fermi-expansion} and \eq{rho-ep} for the ``local'' quantities EE and density  are both governed by a characteristic, locally defined, length-scale given by the lattice spacing \eq{ep-def} (or equivalently, by the short-distance scale \eq{short-dis}). To be precise, the EE of an interval of length $l$, centred around $x$, is of the scaling form
\begin{align}
  S(l, x, g_s, l_s)= f\left(\fr{l}{\ep(x)}\right), \qquad \ep(x)= g_s(x) l_s, \qquad g_s(x)= \fr{g_s}{\sinh^2(x/l_s)}
  \label{ee-scaling}  
  \end{align}
whereas the expectation value of the density $\lan\rho(x)\ran$ is of the form
\begin{align}
  \lan\rho(x)\ran_{_{g_s, l_s}}= \fr1{\ep(x)}.
  \label{rho-scaling}
  \end{align}
Note that even the correction terms in \eqref{subleading-1pt} are given in terms of the length scale $\epsilon(x)$. We also found that this feature does not hold for S-matrix elements such as the two-point function discussed in Section \ref{sec:rho-rho}, which is of the form of a standard string theoretic perturbation expansion
\begin{align}
  G(\om,g_s, l_s)= \sum_{h} g_s^{2-2h} G_h(\om l_s)
  \label{2pt-scaling}
\end{align}

How do we interpret these? In particular, how do we understand the scaling forms \eq{ee-scaling} and \eq{rho-scaling}? Going by the duality between the $c=1$ matrix model and 2D string theory, naively one might expect some form like \eq{2pt-scaling} for all quantities. The hint for the resolution comes from the form of the particular energy scale $\fr1\ep \sim \fr1{g_s l_s}$ which is the order of a D0 brane mass. In the more recent interpretation of the $c=1$ matrix model, the matrix eigenvalues (which are fermionic) are interpreted as the ``locations'' of $N$ unstable D0 branes in a 2D bosonic string theory \cite{McGreevy:2003kb, Martinec:2003ka, Klebanov:2003km, Douglas:2003up, McGreevy:2003dn, Takayanagi:2003sm}.

The ground state of the fermion theory, namely the filled Fermi sea, is identified as the lowest energy configuration of the $N$-soliton state. Low energy fluctuations near the ground state, can be described in terms of massless bosons (``sound modes'') which correspond to the massless scalar of the 2D closed string theory (improperly called a tachyon). Now, both the EE and the expectation value of the density, are calculated in the fermion ground state. They, therefore, belong to the solitonic segment of the theory, which is governed by the length scale given by the inverse soliton mass $\fr1{M_{sol}}=$$ \fr1{1/g_s l_s}= $$ g_s l_s$. Fluctuations appear in the calculation of the two-point function, which, therefore see the string scale.

The above explanation still does not explain the characteristic length scale $\ep(x)$ which is local. A possible way to understand this is to note that the D0 brane at the location $x$ has a binding energy $E(x) = \mu \sinh^2(x/l_s)= {\ep(x)}^{-1}$ relative to the Fermi level. This gives rise to the notion of a position-dependent soliton energy, leading to the position-dependent length scale $\ep(x)$. 

\subsection*{Acknowledgments}
It is a great pleasure to thank Sumit Das, Rajesh Gopakumar, R. Loganayagam, Shiraz Minwalla, Onkar Parrikar, Suvrat Raju, Spenta Wadia and especially Ashoke Sen for crucial discussions. 
One of the authors (G.~M.) would like to acknowledge the Raja Ramanna Chair Professorship.  
G.~M. would also like to thank the organizers of the ``Large $N$ Matrix Models and Emergent Geometry'' BIRS workshop (Banff, May 2026) for an opportunity to present part of this work. A.~M. gratefully acknowledges the hospitality of ICTS during the course of this work and for the opportunity to present a part of this work.

\newpage
\begin{appendix}
\section{Exact bosonization operator mapping}\label{app:op-boson}
The exact operator mapping between the fermion bilinears and the bosonic oscillators are given by
\begin{equation}\label{exact-op-map}
\begin{aligned}
\psi^\dagger_{n+m}\psi_n
={}&
\sum_{k=1}^{N-1}
\Biggl[
\sigma_k^m
(\sigma_{k+1}^\dagger)^m
\theta_+(a_k^\dagger a_k-m)
-\sum_{r_k=0}^{\infty}
\sigma_{k-1}^{m-1-r_k}
(\sigma_k^\dagger)^{m-2-r_k}
\sigma_k^{r_k}
(\sigma_{k+1}^\dagger)^{r_k+1}
\\
&\qquad\times
\theta_-(a_k^\dagger a_k-m+1)
\theta_+\!\left(
a_{k-1}^\dagger a_{k-1}
+a_k^\dagger a_k-m+1
\right)
\delta(a_k^\dagger a_k-r_k)
\\
&\quad
+\sum_{j=2}^{k-1}
(-1)^j
\sum_{r_{k-j+1}=0}^{\infty}
\sum_{r_{k-j+2}=0}^{\infty}
\cdots
\sum_{r_k=0}^{\infty}
\sigma_{k-j}^{
m-j-\sum_{i=1}^{j}r_{k-j+i}}
\\
&\qquad\times
(\sigma_{k-j+1}^\dagger)^{
m-j-1-\sum_{i=1}^{j}r_{k-j+i}}
\sigma_{k-j+1}^{r_{k-j+1}}
(\sigma_{k-j+2}^\dagger)^{r_{k-j+1}}
\cdots
\sigma_{k-1}^{r_{k-1}}
(\sigma_k^\dagger)^{r_{k-1}}
\\
&\qquad\times
\sigma_k^{r_k}
(\sigma_{k+1}^\dagger)^{r_k+1}
\theta_-\!\left(
\sum_{i=1}^{j}
a_{k-j+i}^\dagger a_{k-j+i}
-m+j
\right)
\\
&\qquad\times
\theta_+\!\left(
\sum_{i=0}^{j}
a_{k-j+i}^\dagger a_{k-j+i}
-m+j
\right)
\prod_{i=1}^{j}
\delta\!\left(
a_{k-j+i}^\dagger a_{k-j+i}
-r_{k-j+i}
\right)
\\
&\quad
+(-1)^k
\sum_{r_1=0}^{\infty}
\cdots
\sum_{r_k=0}^{\infty}
(\sigma_1^\dagger)^{
m-1-k-\sum_{i=1}^{k}r_i}
\sigma_1^{r_1}
(\sigma_2^\dagger)^{r_1}
\cdots
\sigma_{k-1}^{r_{k-1}}
(\sigma_k^\dagger)^{r_{k-1}}
\\
&\qquad\times
\sigma_k^{r_k}
(\sigma_{k+1}^\dagger)^{r_k+1}
\theta_-\!\left(
\sum_{i=1}^{k}a_i^\dagger a_i-m+k
\right)
\prod_{i=1}^{k}
\delta(a_i^\dagger a_i-r_i)
\Biggr]
\\
&\quad\times
\delta\!\left(
\sum_{i=k+1}^{N}a_i^\dagger a_i
-n+N-k-1
\right)
\\
&\quad+
(\sigma_1^\dagger)^m
\delta\!\left(
\sum_{i=1}^{N}a_i^\dagger a_i
-n+N-1
\right)
\end{aligned}
\end{equation}
where
\begin{align}
\sigma_k = \frac{1}{\sqrt{a_k^{\dagger} a_k +1}} a_k \qquad \sigma_k^{\dagger} = a_k^{\dagger} \frac{1}{\sqrt{a_k^{\dagger}a_k+1}}
\end{align}
The above expression is only valid for $m>0$ and the $m<0$ case can be obtained from it by conjugation. The inverse relations are given by
\begin{equation}\label{fermionize1}
\begin{aligned}
\ad_k = & \sum_{m_k> m_{k-1} > \cdots > m_0}
\sqrt{m_1 - m_0}~(\psid_{m_0}\psi_{m_0})(\psid_{m_1 + 1} \psi_{m_1}) \cdots
(\psid_{m_k + 1} \psi_{m_k})  \\
& ~~\times \delta \biggl (\sum_{m=m_0 +1}^{m_1 -1} 
\psid_m \psi_m \biggr )~
\delta \biggl (\sum_{m=m_1 +1}^{m_2 -1}\psid_m \psi_m \biggr ) \cdots
~\delta \biggl (\sum_{m=m_{k-1} +1}^{m_k -1} \psid_m \psi_m \biggr )  \\
& ~~\times 
\delta \biggl (\sum_{m=m_k +1}^{\infty}\psid_m \psi_m \biggr ), 
\quad \quad \quad \quad k=1, 2, \cdots , (N-1)
\end{aligned}
\end{equation}
and
\begin{equation}\label{fermionize2}
\begin{aligned}
\ad_N = & \sum_{m_N> m_{N-1} > \cdots > m_1}
\sqrt{m_1 + 1}~(\psid_{m_1 + 1} \psi_{m_1}) \cdots 
(\psid_{m_N + 1} \psi_{m_N})  \\
& ~~~~~~~~~~~~\times 
\delta \biggl (\sum_{m=m_1 +1}^{m_2 -1}\psid_m \psi_m \biggr ) \cdots
~\delta \biggl (\sum_{m=m_{N-1} +1}^{m_N -1} \psid_m \psi_m \biggr )  \\
& ~~~~~~~~~~~~\times 
\delta \biggl (\sum_{m=m_N +1}^{\infty}\psid_m \psi_m \biggr ).
\end{aligned}
\end{equation}
The expressions for the annihilation operators $a_k$, $k=1,2,...,N$ can be obtained by taking the adjoint of the above equations.

\section{Relation between the lattice boson and the phase space distribution}\label{app:lattice-vs-uxp}
The eigenvalue density is not sufficient to describe MQM (and its equivalent fermionic system) completely. It holds specifically in the approximation where the Wigner density is a droplet in phase space and where no shock formation or fold development has taken place in the droplet. To completely characterize the fermionic system, we must consider the Wigner phase space distribution.  
It is determined by the fermionic bilinears in the following way
\begin{align}
\hat{u}(\lambda,p) = \int d\eta\; \psi^{\dagger}(x+\eta) \psi(x-\eta) e^{i 2 p \eta} = \sum_{m,n=0}^{\infty}\psi^{\dagger}_m \psi_n \int d\eta \; \psi^*(x+\eta) \psi(x-\eta) e^{i 2p \eta}
\end{align}
Given that the bilinears $\psi^{\dagger}_m \psi_n$ can be mapped to the bosonic oscillators ($a_k$ and $a^{\dagger}_k$) which can in turn be mapped to the local bosonic fields $\phi(x_j)$ and $\pi(x_j)$, we can immediately see that the Wigner distribution can be mapped to our lattice boson fields $\phi(x_j)$ and $\pi(x_j)$, and vice-versa. 

Moreover, the operator-level constraints of the fermionic theory
\begin{align}
\int dy\;  \tilde \Phi(x,y) \tilde \Phi(y,z) = N \tilde \Phi(x,z), \qquad \tilde \Phi(x,y) \equiv \Psi(x) \Psi^{\dagger}(y)
\end{align}
are trivially satisfied by our lattice boson operators ($\phi(x_j)$ and $\pi(x_j)$), thereby providing an unconstrained bosonic description of the fermionic theory. 


Since the relation between the closed-string tachyon and the Wigner distribution is established in \cite{Dhar:1995np}, we may therefore in principle relate our lattice boson directly to the string theory tachyon. One may wonder what happens to the UV-cutoff $\mu$ of the lattice boson theory (see, e.g. \eq{lattice-ham}) when one maps to the string theory tachyon; the collective field theory action \eq{collective-action}, which is related by an integral transform to the classical action of the closed string tachyon, does not appear to have any lattice cut-off or a UV-cutoff. The reason for this is that the map in \cite{Dhar:1995np} is perturbative, valid at low energies; a more accurate map would transform the Wigner distribution to a string field whose action will have a UV-cutoff at the string scale automatically. With such a map, one would be able to understand the finite results in \eq{2pt-scaling} automatically.\footnote{The finite results \eq{ee-scaling} and \eq{rho-scaling} would require a map of the lattice boson theory to a theory of D0-branes; in principle, this is simply the relation between the lattice boson and the fermions.}

\section{Derivation of the collective field theory action} \label{app:collective-action-derivation}


We parameterize the fermi fluid by the following phase space density
\begin{align}
  u(\l, p,t)= \theta(p_+(\l,t)- p)\theta(p- p_-(\l,t))
\label{u-pq}
\end{align}
The single-particle Hamiltonian
\begin{align}
  h(\l,p)= \fr{p^2}{2}+ V(\l)
  \label{h-pq}
  \end{align}
We begin with the Hamiltonian in terms of $p_+(\lambda)$ and $p_-(\lambda)$
\begin{align}
  H &=\frac{1}{2\pi} \int d\lambda\,dp\, u(\l,p,t)h(\l,p)
  \nonumber\\
  &= \frac{1}{2\pi} \int d\lambda\, \left[ \frac{1}{6} (p_+^3 - p_-^3) +V(\l) (p_+ - p_-) \right]
\end{align}
where $p_{\pm}(\lambda)$ satisfy the following Poisson bracket relations
\begin{align}
\{p_{\pm}(\lambda), p_{\pm}(\lambda')\} = - 2\pi  \partial_{\lambda} \delta(\lambda - \lambda'), \qquad \{p_+(\lambda), p_-(\lambda')\} = 0 
\end{align}
We express $p_{\pm} (\lambda)$ in terms of new fields $\Pi_\varphi(\lambda)$ and $\varphi(\lambda)$ that satisfy the usual Poisson bracket  
\begin{align}
p_{\pm}(\lambda) = - \Pi_\varphi(\lambda) \pm \pi \partial_{\lambda}\varphi(\lambda), \quad \{\varphi(\l), \Pi_\varphi(\l')\}= \delta(\l  -\l') 
\end{align}
The Hamiltonian expressed in these new variables becomes
\begin{align}
  H = \int d\lambda \, \left[ \frac{1}{2} \Pi_\varphi^2 \varphi ' + \frac{\pi^2}{6} \varphi'^3 + V(\l) \varphi'\right].
  \label{coll-ham}
\end{align}
The action then is given by
\begin{align}
S = \int dt \,d\lambda\, \left[ \Pi_\varphi \dot\varphi - \frac{1}{2} \Pi_\varphi^2 \varphi' - \frac{\pi^2}{6} \varphi'^3 + V(\l) \varphi' \right], \qquad \varphi' \equiv \partial_{\lambda}\varphi \;\& \; \dot\varphi \equiv \partial_t\varphi
\label{coll-action1-app}
\end{align}
Using the equation of motion of $\Pi_\varphi$ ($
\Pi_\varphi = \dot\varphi/\varphi'$), we can write the action completely in terms of the $\varphi$ variable as follows
\begin{align}
S = \int dt \, d\lambda \, \left[ \frac{1}{2} \frac{\dot\varphi^2}{\varphi'} - \frac{\pi^2}{6} \varphi'^3 + \frac{\lambda^2}{2} \varphi' - \mu \varphi' \right]
\label{coll-action2-app}
\end{align}
where we have chosen the inverted harmonic oscillator $V(\l)= -\l^2/2$ and added a the contribution of the Fermi energy $\mu$. Since we are interested in fluctuations around the ground state, we will now perform a background subtraction of the Fermi sea
\begin{align}
\varphi(\lambda,t) = \varphi_0(\lambda) + \phi(\lambda,t), \qquad \partial_\lambda \varphi_0 =\rho_0 = \frac{1}{\pi}\sqrt{\lambda^2 - 2 \mu} \label{fluct}
\end{align}
to write our theory in terms of the fluctuation variable $\phi(\lambda,t)$
\begin{align}
  S = \int dt\, d\lambda \, \left[ \frac{1}{2} \frac{\dot \phi^2}{\rho_0 + \phi'} - \frac{\pi^2}{2} \rho_0 \phi'^2 - \frac{\pi^2}{6} \phi'^3 + \frac{\pi^2}{3} \rho_0^3 \right]
\label{coll-action3-app}
\end{align}
The last constant term will be omitted. Next we use the following map to string theory variables
\begin{align}
\lambda = \sqrt{2 \mu} \cosh x, \qquad g_s = \frac{1}{\mu}
\end{align}
to re-express the action as
\begin{align}
  S = \frac{\pi}{2} \int dt\,dx\, \left[ \dot\phi^2\left(1+\frac{\pi g_s}{2\sinh^2x}\phi'\right)^{-1} - \phi'^2 - \frac{\pi g_s}{6 \sinh^2x} \phi'^3 \right], \qquad \phi' \equiv \partial_x \phi \; \& \; \dot \phi \equiv \partial_t \phi
  \label{coll-action4-app}
\end{align}
Expanding to the second order in $g_s$ 
\begin{align}
S = \frac{\pi}{2} \int dt\, dx\, \left[\dot\phi^2 - \phi'^2 - \frac{\pi g_s}{6 \sinh^2x} (3 \dot\phi^2 \phi' - \phi'^3) + \frac{\pi^2 g_s^2}{4 \sinh^4 x} \dot\phi^2 \phi'^2 \right] + \mathcal{O}(g_s^3)
\end{align}
and expressing in terms of the lightcone coordinates $\del_{\pm} = (\del_t \pm \del_x)/2$, we get our final result
\begin{align}
  \scalebox{0.85}{$\displaystyle S =2 \pi \int dz^+ dz^- \left[ \partial_+\phi \partial_- \phi - \frac{\pi g_s}{6 \sinh^2 x} \left( (\partial_+ \phi)^3 - (\partial_-\phi)^3\right) + \frac{\pi^2 g_s^2}{4 \sinh^4x} \left( (\partial_+ \phi)^2 - (\partial_-\phi)^2 \right)^2 \right] +\mathcal{O}(g_s^3)$}
  \label{coll-action5-app}
\end{align}
We identify $\rho(\lambda)$ with $\partial_{\lambda} \varphi$, which implies the fluctuation of the density in terms of $\phi$ is given by
\begin{gather}
\rho = \rho_0 + \delta \rho = \partial_{\lambda} \varphi_0 + \partial_{\lambda} \phi \\
\Rightarrow \delta \rho = \partial_{\lambda} \phi =\left(\frac{d \lambda}{dx}\right)^{-1} \partial_x \phi  = \frac{1}{\rho_0(\lambda)} \partial_x \phi
\end{gather}

\section{Cubic interaction $1 \to 1$ scattering amplitude from collective field theory}
\label{app:rho-rho}
Unlike the main body of the paper, the computation in this section works with $f(x) = {\pi}/{(6 \sinh^2x)}$ everywhere instead of $g_s(x) = g_s/\sinh^2(x)$. The main purpose of this section is to elaborate the steps on the momentum integrals in light-cone coordinates. The collective field theory action (including just the cubic interaction) is given by
\begin{align}
\scalebox{1}{$\displaystyle
S[\phi]$} &\scalebox{1}{$\displaystyle = 2\pi \int dt d\tau \left(\partial_+\phi \partial_-\phi - g_{s}\ f(\tau) \left[(\partial_+\phi)^3 - (\partial_-\phi )^3\right]\right)$} \\
&\scalebox{0.75}{$\displaystyle = -2 \pi^3 \left( \int dp^2\ \tilde\phi(p)\ (k^2 - \omega^2)\ \tilde\phi(-p) - \f{i\ g_s}{2} \int d \tilde k dp^2 dp'^2 \tilde f(\tilde k) \tilde \phi(p)\tilde \phi(p')\tilde \phi(p'') \left[(p^-)(p'^-)(p''^-) + (p^+)(p'^+) (p''^+) \right] \right)$}
\end{align}
The cubic contribution is then given by
\begin{equation}
\resizebox{0.9 \textwidth}{!}{$ \displaystyle I_{\rm cubic} = -C_1 g_s^2 \pi^6 \left(\f{1}{4\pi^2}\right)^2  p_1^+ p_2^+ \int dk d\omega_T dk_T\ \f{f(k)^2\ (p_T^+)^2(\tilde{p}_T^+)^2}{(\omega_T^2 - k_T^2 + i \ep)((\omega_1 - \omega_T)^2 - (k_1 + k - k_T)^2 + i \ep)}  $}
\end{equation}
Let us redefine \(\omega = \omega_1 = \omega_2 = k_1 = k_2\) then
\begin{equation}
I_{\rm cubic}= -\f{C_1 g_s^2 \pi^2}{4}\ \omega^2 \int dk d\omega_T dk_T\ \f{f(k)^2\ (p_T^+)^2(\tilde{p}_T^+)^2}{(\omega_T^2 - k_T^2 + i \ep)((\omega - \omega_T)^2 - (\omega + k - k_T)^2 + i \ep)}   
\end{equation} 
Let us also redefine \(p_T =p\) and we get  
\begin{equation}
\resizebox{0.9\textwidth}{!}{$\displaystyle I_{\rm cubic} = -\f{C_1 g_s^2 \pi^2}{8}\ \omega^2 \int dk dp^+\ f(k)^2\ (p^+)^2(2\omega + k - p^+)^2 \int  dp^-\ \f{1}{(p^+p^- + i \ep)[(k-p^+)(2\omega - k - p^-)+i\ep]}   $}
\end{equation} 
rearranging this we get 
\begin{equation}
\resizebox{0.9\textwidth}{!}{$\displaystyle 
I_{\rm cubic} = -\f{C_1 g_s^2 \pi^2}{8}\ \omega^2 \int dk dp^+\ f(k)^2\ (p^+)^2(2\omega + k - p^+)^2 \int  \f{ dp^-}{(p^+p^- + i \ep)[p^- (p^+ - k -2\omega) + k  (p^+ - k -2\omega) + i\ep]}   $}
\end{equation}  
we can do the \(p^-\) integrals and it limits the \(p^+\) integrals 
\begin{equation}
I_{\rm cubic} = -\f{C_1 g_s^2 \pi^2}{8}\ \omega^2 \int dk dp^+\ f(k)^2\ (p^+)^2(2\omega + k - p^+)^2 \hspace{6cm} \nonumber \\ \hspace{2cm} 2\pi\left(\underbrace{\f{\theta(-p^+)\theta( p^+ - k - 2\omega)}{i p^+ k (p^+ - k - 2\omega) + \ep (k + 2\omega)}}_{(k+2\omega) < p^+ < 0} - \underbrace{\f{\theta(p^+)\theta(k + 2\omega - p^+)}{i p^+ k (p^+ - k - 2\omega) + \ep (k + 2\omega)}}_{0 > p^+ > (k + 2\omega)}\right). 
\end{equation}  
We can rewrite above integral as 
\begin{equation}
I_{\rm cubic}= -\f{C_1 g_s^2 \pi^3}{4}\ \omega^2 \int dk f(k)^2 \left(\int_{k+2\omega}^0 dp^+ \f{(p^+)^2(2\omega + k - p^+)^2}{i p^+ k  (p^+ - k - 2\omega) + \ep (k + 2\omega)} \hspace{6cm} \right. \nonumber \\ \left. \hspace{6cm}- \int_0^{k + 2\omega} dp^+\f{(p^+)^2(2\omega + k - p^+)^2}{i p^+ k  (p^+ - k - 2\omega) + \ep (k + 2\omega)}\right)  
\end{equation} 
lets set \(\ep \to 0\)
\begin{equation}
\resizebox{0.9\textwidth}{!}{$\displaystyle I_{\rm cubic} = -i\f{C_1 g_s^2 \pi^3}{4}\ \omega^2 \int \f{dk}{k} f(k)^2 \left(\int_{k+2\omega}^0 dp^+\ p^+\ (2\omega + k - p^+) - \int_0^{k + 2\omega} dp^+\ p^+\ (2\omega + k - p^+) \right)  $}
\end{equation} 
integrating \(p^+\) we get 
\begin{equation}
I_{\rm cubic} = i\f{C_1 g_s^2 \pi^3}{24}\ \omega^2 \int \f{dk}{k} f(k)^2\ (k + 2\omega)^3.
\end{equation}
We now set \(f(\tau) =  {\pi}/{6 \sinh(\tau)^2}\) and \(f(k) = -(2 + k\pi \coth (k\pi/2))/12\)
\begin{equation}
I_{\rm cubic} = i\f{C_1 g_s^2 \pi^3}{3456}\ \omega^2 \int dk\ \left(2 +  k\pi\ \coth \left(\f{k\pi}{2}\right) \right)^2\ \left(k^2 + 6k \omega + 12 \omega^2 + \f{8\omega^3}{k}\right)
\end{equation}
The integration is UV divergent so we put a UV cutoff \(k \in (-\Lambda, \Lambda)\), this gives us 
\begin{equation}
\resizebox{0.9\textwidth}{!}{$\displaystyle I_{\rm cubic} = i\f{C_1 g_s^2 \pi^3}{3456}\  \left[\omega^2 \left(\f{16 \pi}{5} + \f{8\Lambda^3}{3} + 2\pi \Lambda^4 + \f{2\pi^2 \Lambda^5}{5}\right) + \omega^4 \left(64 \pi + 96 \Lambda + 48 \pi \Lambda^2  + 8 \pi ^2 \Lambda ^3  \right) - i\ \omega^5\ 128\pi    \right] $}
\end{equation}
rearranging above expression
\begin{equation}
\resizebox{0.9\textwidth}{!}{$\displaystyle I_{\rm cubic} = i\f{C_1 g_s^2 \pi^4}{1080}\  \left[\omega^2 \left(\frac{\pi  \Lambda ^5}{8}+\frac{5 \Lambda ^4}{8}+\frac{5 \Lambda ^3}{6 \pi } + 1\right) + \omega^4 \left(\frac{5 \pi  \Lambda ^3}{2}+15 \Lambda ^2+\frac{30 \Lambda }{\pi }+20  \right) - i\ \omega^5\ 40   \right]$}
\end{equation}

\section{Keeping track of $l_s$ factors}\label{app:ls-factor}
In this section, we derive the action of the double-scaled $c=1$ matrix model from the BFSS matrix model, keeping a careful account of all the string length $l_s$ factors. The Dirac-Born-Infeld (DBI) action a single BPS D0 brane in 2D is
\begin{align}
S_{\rm BPS} = \frac{1}{g_s l_s} \int dt\; \Tr \sqrt{1 - \dot X^2}
\end{align}
 which informs us that the correct factors of $l_s$ in the action for a non-BPS D0 brane (provided in \cite{McGreevy:2003kb}) is
 \begin{align}
 S_{\rm non-BPS} &
 \approx \frac{N}{\lambda l_s} \int dt \; \Tr \left( \frac{\dot T^2}{2} + \frac{T^2}{2 l_s^2} - \frac{T^4}{4 l_s^4} \right)
 \end{align}
 If we define $T/l_s$ as the matrix $M$, we get the action for the $c=1$ matrix model before the double-scaling limit is taken.
 \begin{align}
 S &= \frac{N \beta}{ l_s} \int dt \; \left(\frac{l_s^2}{2} \dot M^2 - V(M)\right), \qquad V(M) = -\frac{1}{2}M^2 + \frac{1}{4}M^4 \\
 &= \frac{N \beta}{l_s} \int dt \; \sum_{i=1}^N\left(\frac{l_s^2}{2} \dot \lambda_i - V(\lambda_i)\right)
 \end{align}
 The single-particle Hamiltonian is then given by
 \begin{align}
 h &= \frac{1}{l_s} \left( \frac{\hbar^2}{\beta N} \frac{\partial}{\partial \lambda^2} -  \lambda^2 + \beta N \lambda^4 \right) 
 \end{align}

\end{appendix}

\bibliography{references.bib}
\bibliographystyle{JHEP} 

\end{document}